\documentclass[fleqn,usenatbib,letters]{mnras}
\usepackage[british]{babel}             
\let\la=\lesssim
\usepackage[T1]{fontenc}                
\usepackage{ae,aecompl}
\usepackage{graphicx}	
\usepackage{amsmath}	
\usepackage{amssymb}	
\newcommand\mnraslet{MNRAS}             
\newcommand\ijmpd{Int. J. Mod. Phys. D}     
\newcommand\pa{Phys. Rev.}     

\title[Non-zero active mass in FRW models] {Friedmann--Robertson--Walker models do not require zero active mass}
\author[Do Young Kim et al.]{
Do Young Kim,$^{1,2}$\thanks{dyk25@mrao.cam.ac.uk}
Anthony N. Lasenby,$^{1,2}$\thanks{a.n.lasenby@mrao.cam.ac.uk}
and Michael P. Hobson$^{1}$\thanks{mph@mrao.cam.ac.uk}
\\
$^{1}$Astrophysics Group, Cavendish Laboratory, JJ Thomson Avenue, Cambridge CB3 0HE\\
$^{2}$Kavli Institute for Cosmology, Madingley Road, Cambridge CB3 0HA\\
}

\date{Accepted???. Received???; in original form 27 January 2016}

\pubyear{2016}

\begin{document}
\label{firstpage}
\pagerange{\pageref{firstpage}--\pageref{lastpage}}
\maketitle

\begin{abstract}
The $R_{\rm h}=ct$ cosmological model has received considerable
attention in recent years owing to claims that it is favoured over the
standard $\Lambda$ cold dark mater ($\Lambda$CDM) model by most observational data. A key feature
of the $R_{\rm h}=ct$ model is that the zero active mass condition
$\rho+3p=0$ holds at all epochs. Most recently, \cite{Melia2016} has claimed that
this condition is a requirement of the symmetries of the
Friedmann--Robertson--Walker (FRW) spacetime. We demonstrate that this claim
is false and results from a flaw in the logic of Melia's argument.
\end{abstract}

\begin{keywords}
cosmology: theory -- gravitation
\end{keywords}



\section{Introduction}

The $\Lambda$CDM model serves as the basis for the current standard
model of cosmology, which provides a good fit to a wide range of
cosmological observations. As pointed out by~\cite{Melia2003},
however, for the best-fit $\Lambda$CDM model, the present-day Hubble
distance is broadly consistent with $ct_0$ to within observational
uncertainties, where $t_0$ is the current cosmic epoch. In other
words, observations suggest that the universe has expanded by an
amount similar to what would have occurred had the expansion rate been
constant or, equivalently, that the average acceleration of the
universe up to the present epoch is consistent with zero; this is despite
the fact that the combination of time-dependent radiation, matter and
dark-energy densities $\rho_{\rm r}(t)$, $\rho_{\rm m}(t)$ and
$\rho_{\rm de}(t)$ should have produced periods of deceleration and
acceleration.  Another way to describe this finding \citep{Melia2009}
is that, averaged over a Hubble time, the quantity $p/\rho$, where
$\rho = \rho_{\rm r} + \rho_{\rm m} + \rho_{\rm de}$ and $p = p_{\rm
  r} + p_{\rm m} + p_{\rm de}$, yields $\langle p/\rho\rangle = -1/3$
to within the observational uncertainties.

In the $\Lambda$CDM model, this correspondence is a peculiar
coincidence, made all the more striking by the fact that, for the
best-fit model, this situation should occur only once in the history
of the universe. The fact that we observe this correspondence at the
present epoch is therefore most intriguing.

Consequently, it was proposed
by~\cite{Melia2007,Melia2009,MeliaShevchuk2012} that this
correspondence is not coincidental, but should be satisfied at all
cosmic times $t$. The physical argument originally presented for this
viewpoint was based on applying Birkhoff's theorem and its corollary
to a spherical subregion of a homogeneous and isotropic matter
distribution, from which it was claimed that one could identify a
gravitational radius $R_{\rm h}=2GM/c^2$, given in terms of the
Misner--Sharp mass $M = (4\pi/3)R_{\rm h}^3(\rho/c^2)$
\citep{MisnerSharp1964}. Moreover, it is easily shown that $R_{\rm h}$
coincides with the Hubble radius in a spatially-flat universe containing any single fluid component. 
In particular, it was claimed that
Weyl's postulate requires $R_{\rm h}$, and hence the Hubble radius, to
be a `proper' distance, i.e. one that is comoving with the cosmological
fluid. Imposing this condition on the usual cosmological field
equations for an FRW spacetime picks out a unique solution, for which
$R_{\rm h}(t)=ct$ at all cosmic times. This is equivalent to vanishing
total active mass, $\rho+3p=0$, at all epochs. The resulting
cosmological model, known as the `$R_{\rm h}=ct$' model, has received
considerable attention over the last few years, since it has been
claimed to be favoured over the standard $\Lambda$CDM (and its variant
$w$CDM with $w\neq -1$) by most observational data
\citep{MeliaMaier2013,Wei2013,Wei2014a,Wei2014b,Wei2015,Melia_etal2015}.

Recent observational data have, however, led to serious criticisms of
the $R_{\rm h}=ct$ model. For example, the model requires the
deceleration parameter $q(z)=0$ at all times, but current data from
supernovae and baryon acoustic oscillations strongly disagrees with
$q_0=0$ at high significance \citep{Bilicki2012}, and robust model
comparison methods strongly disfavour the $R_{\rm h}=ct$ model
\citep{Shafer2015}.  In addition, recent cosmic microwave background (CMB) data from the
\textit{Planck} satellite rule out the equivalence of the age of the
universe to $1/H_0$ at greater than 99 per cent confidence, favouring
$R_{\rm h} = \left(1.05 \pm 0.02\right)ct$ at the current epoch
\citep{vanOirschot2014}, which undermines a major motivation for the
$R_{\rm h}=ct$ model; note that this result is equivalent to $q_0=0.05
\pm 0.02$ \citep{vanOirschot2014}.

In addition to objections based on observations, the validity
of the theoretical argument underlying the $R_{\rm h}=ct$ model has also
been criticised by a number of authors
\citep{vanOirschot2010,LewisvanOirschot2012,Mitra2014}, and in
particular the validity of the effective equation-of-state parameter
$w=-1/3$ \citep{Lewis2013}.  These and other criticisms are claimed to
have been addressed by~\cite{Bikwa2012} and~\cite{Melia2012c} (see
also~\cite{Melia2015} and references therein), but the original
physical arguments for the model given
in~\cite{Melia2007,Melia2009,MeliaShevchuk2012} are sufficiently
imponderable that it is difficult to draw definite conclusions.

In a recent paper \citep{Melia2016}, however, Melia presents a much
more explicit argument for the zero active mass condition $\rho+3p=0$,
which he claims is a requirement of the symmetries of the FRW
spacetime. In particular, it is claimed that assuming the general,
spherically symmetric (but radially varying) metric, solving the
Einstein field equations, and then imposing homogeneity and isotropy
yields an extra condition, namely vanishing active mass, which is lost
if one adopts the usual procedure of first imposing the conditions of
homogeneity and isotropy on the metric and then solving the Einstein
equations. The purpose of our Letter is to demonstrate that this claim
is false, owing to a flaw in the logic of Melia's argument, and hence
that FRW models are in fact consistent with non-zero active mass.

\section{FRW metric and zero active mass}

Melia starts with the general spherically symmetric metric in a
comoving coordinate system, which we denote by
\begin{equation}
  ds^2 = A^2dt^2 - B^2dr^2 - R^2 d\Omega^2,
  \label{eqn:metric_comov}
\end{equation}
where $A$, $B$ and $R$ are in general functions of both $r$ and $t$, and first considers the general case, where homogeneity is \textit{not} assumed. Using the Einstein equations, assuming zero cosmological constant, one may derive the Euler and continuity equations
\begin{align}
  \frac{\partial p}{\partial r} & =  -\frac{1}{A}\frac{\partial
    A}{\partial r} (\rho + p),  \label{eqn:dpdr_comov} \\
\dot{\rho} & =  -\left( \rho + p \right) \left( 2\frac{\dot{R}}{R}   + \frac{\dot{B}}{B}  \right),
  \label{eqn:cont} 
\end{align}
where a dot denotes differentiation with respect to $t$. Incidentally,
in his equation (13), Melia gives an incorrect form of the continuity equation, $\dot{\rho}  = -3\left( \rho + p \right) (\dot{R}/R)$, which is valid only in the homogeneous case, but this error has no bearing
on the rest of his argument.

Melia then imposes homogeneity and finds from equation~\eqref{eqn:dpdr_comov} that $A$ is independent of $r$, such that $A=A(t)$. Moreover, as usual, one may also write $B(r,t)=a(t)/\sqrt{1-kr^2}$ and $R(r,t)=a(t)r$, where $a(t)$ is the scale factor and $k$ is the spatial curvature constant. The Einstein equations then yield the corresponding Friedmann equation and acceleration equation,
\begin{align}
    {\left( \frac{\dot{a}}{a} \right)}^2 &= \frac{8\pi}{3}\rho A^2 - \frac{k}{a^2}A^2,
    \label{eqn:Friedmann_melia}\\
    \frac{\ddot{a}}{a} - \frac{\dot{a}}{a} \frac{d\ln A}{dt} &= -\frac{4\pi}{3} A^2 (\rho + 3p),
    \label{eqn:accn_melia}
\end{align}
which Melia combines into the single equation
\begin{equation}
    \frac{d}{d t} \left[ \ln\left( \frac{\dot{a}^2}{A^2} \right) \right] =- \left( \frac{k}{a\dot{a}} A^2 +  \frac{\dot{a}}{a} \right) \left( 1+ \frac{3p}{\rho} \right).
    \label{eqn:melia_28}
\end{equation}

Melia then writes $A(t)$ in the following form
\begin{equation}
    A^2(t) = h\dot{a}^2 e^{I(t)},
    \label{eqn:test_form_for_A}
\end{equation}
where $h$ is a constant and $I(t)$ is a function defined by the
above equation.  Substituting (\ref{eqn:test_form_for_A}) into the LHS of~\eqref{eqn:melia_28} and using~\eqref{eqn:Friedmann_melia}, then gives
\begin{equation}
    \frac{d I(t)}{d t}= \frac{8\pi}{3}\frac{a}{\dot{a}}A^2 \left( \rho+3p \right).
    \label{eqn:form_for_I}
\end{equation}

The flaw in his logic then lies in the following. He asserts that, in
order for $A$ to be a constant, as it is in the FRW metric, 
equation~\eqref{eqn:test_form_for_A} requires 
\textit{both} $\dot{a}^2$ and $e^{I(t)}$ to be constant in
time. This \emph{incorrect}
assertion then leads one to conclude that $dI(t)/dt=0$ at
all times, and that by equation~\eqref{eqn:form_for_I}, $\rho+3p=0$ at
all times. He therefore concludes that the FRW metric (for which
$A=1$) requires the zero active mass condition to be satisfied.  This
assumption is clearly wrong, however, as the RHS of
equation~\eqref{eqn:test_form_for_A} can be constant without
$\dot{a}^2$ and $e^{I(t)}$ both being constant.

That equations \eqref{eqn:test_form_for_A} and
\eqref{eqn:form_for_I} can be satisfied for $A=\mbox{constant}$ and
$\rho+3p\neq 0$ is easily illustrated by a simple example.  Let us
consider the conventional FRW metric, for which $A=1$, and
specifically the EdS model (which Melia himself uses as an example to
support his theory), for which $a(t) \propto t^{2/3}$,
$\rho(t)=1/(6\pi t^2)$ and $p(t)=0$. We can use these expressions to
evaluate the RHS of equation~\eqref{eqn:form_for_I} and integrate to
find that $e^{I(t)} \propto t^{2/3}$. Since $\dot{a} \propto
t^{-1/3}$, the powers of $t$ cancel out on the RHS
of~\eqref{eqn:test_form_for_A}, showing that $A$ is a constant, as
required. It is worth noting that in the above analysis, we have {\em
  not} simply imposed $A=1$ {\em a priori}, as in the usual procedure
for deriving the cosmological field equations, but instead
demonstrated that equations \eqref{eqn:test_form_for_A} and
\eqref{eqn:form_for_I}, derived by solving the Einstein equations for
the general spherically-symmetric metric \eqref{eqn:metric_comov},
admit solutions for which $A$ is constant and $\rho+3p\neq 0$.  This
counter-example alone thus disproves Melia's central claim.


It is worth making a few further points regarding his argument for
zero active mass before moving on to the next part of his
argument. First, the expression~\eqref{eqn:form_for_I} that Melia
presents is strange in that it contains $A$, which one may eliminate
in favour of $I(t)$ using~\eqref{eqn:test_form_for_A}. In fact, one
can derive the following two expressions for $I(t)$,
\begin{align}
e^{I(t)} & =  \frac{3}{h(8\pi\rho a^2-3k)}, \label{ei1}\\
e^{-I(t)} & =  -\frac{8\pi}{3}h\int_0^t a\dot{a}(\rho+3p)\,dt', \label{ei2}
\end{align}
%
which make no explicit reference to $A$. Given forms for $\rho$ and $a$ as functions of $t$, we can use either of these equations to compute $I(t)$, and then by using equation~\eqref{eqn:test_form_for_A} can find the $A(t)$ implied. We adopt this route in the example studied in the next section. Alternatively, one could start from a fixed form of $A$, and work forwards from there.  For example, if $A = 1$, then \eqref{eqn:Friedmann_melia} and \eqref{eqn:accn_melia} reduce to the conventional cosmological field equations, and for any solution of them (i.e. for any standard cosmological model) either of the expressions~\eqref{ei1} or~\eqref{ei2} provides an explicit expression for $I(t)$, which when substituted into~\eqref{eqn:test_form_for_A} yields unity on the LHS. Alternatively, if $A$ is not equal to unity, then the solution for $a$ of~\eqref{eqn:melia_28} will differ from that obtained from the usual cosmological field equations, for which $A = 1$, but this would result in a different expression for $I(t)$, sufficient to combine with $\dot{a}^2$ in~\eqref{eqn:test_form_for_A} to recover the corresponding expression for $A$ on the LHS. 

\section{Comoving and free-fall frames}

Having shown above that having $A=\mbox{constant}$ in 
\eqref{eqn:metric_comov} does not require
zero active mass, we now address the second part of Melia's argument,
in which he claims to provide a justification for requiring $A$ to
equal unity; this claim is also incorrect.

In the coordinates defined by~\eqref{eqn:metric_comov}, he first shows
that the 4-velocity of an observer comoving with the fluid is
\begin{equation}
    u^0 = 1/A, \qquad u^i=0 \quad (i=1,2,3),
    \label{eqn:4_vel}
\end{equation}
where the condition $u^i=0$ shows that $r$, $\theta$ and $\phi$ are
comoving coordinates.  He then points out correctly that a
free-falling observer is comoving with the fluid, but goes on to
suggest incorrectly that this implies that the proper time of a
comoving observer must equal the coordinate time $t$ and hence that
$A=1$. He further notes that if $A$ were a function of $t$ (which,
according to his incorrect reasoning addressed above, would be
necessary if $\rho+3p \neq 0$), one might attempt to perform a gauge
transformation of the form
\begin{equation}
    d \tilde{t}=Adt,
    \label{eqn:melia_gauge_transf}
\end{equation}
which would reduce the metric back to the FRW form, with $g_{\tilde{t}
  \tilde{t}}=1$, but he claims that this is not permitted because of
the uniqueness of the comoving, free-falling frame.

These claims are easily demonstrated to be false. As we show below,
the coordinate time $t$ is allowed to be any function of the proper
time of a comoving observer, $\tau$. Therefore, a gauge transformation
of the form given by~\eqref{eqn:melia_gauge_transf} \textit{is}
allowed, and hence it is possible to have $A$ to be dependent on $t$
without any problems.


To illustrate this explicitly, let us consider a cosmology for which
the evolution of the scale factor as a function of coordinate time $t$
is that in Melia's own model, namely $a(t)=bt$, where $b$ is a
constant. Moreover, again following Melia, we will assume that
$k=0=\Lambda$, but instead of his assumption concerning $p= -\frac{1}{3} \rho$, we take the cosmic fluid to have zero
pressure, so that the underlying {\em physical} cosmology is the
Einstein--de--Sitter (EdS) model.

Substituting $a(t)=bt$ into the continuity equation \eqref{eqn:cont}
for the case of a homogeneous universe, one finds
\begin{equation}
    \rho = \frac{C}{t^3},
    \label{eqn:rho_ex1}
\end{equation}
where $C$ is a constant. Substituting this form for the density into our
equation for $e^{I(t)}$ in~\eqref{ei1}, we find that for $k=0$,
\begin{equation}
    e^{I(t)} = \frac{3t}{8\pi h Cb^2},
    \label{eqn:ei_ex1}
\end{equation}
and using equation~\eqref{eqn:test_form_for_A}, we find that
\begin{equation}
    A(t)=\sqrt{\frac{3t}{8\pi C}},
    \label{eqn:A_ex1}
\end{equation}
which is clearly not constant. 

It is straightforward to find the rest
of the metric components, which are
\begin{align}
    B(r,t) & =  f'(r)bt, \label{eqn:B_ex1} \\
    R(r,t) & =  f(r)a(t) = f(r)bt , \label{eqn:R_ex1} 
\end{align}
where $f'(r)=df/dr$ and $f(r)$ is some function of $r$. With these
expressions for $A$, $B$ and $R$, we can use the Einstein equations to
determine the corresponding stress-energy tensor of the cosmic
fluid. As expected, it yields a fluid of density $\rho = C/t^3$, zero
pressure, and 4-velocity given by
\begin{equation}
    u^0 = 1/A =\sqrt{\frac{8\pi C}{3t}} , \qquad u^i=0 \quad (i=1,2,3).
    \label{eqn:4_vel_ex1}
\end{equation}
This shows that we are in a frame comoving with the fluid, but there
is no requirement that the proper time of a comoving observer must
coincide with the coordinate time $t$.

Indeed, we can find the explicit relationship between $t$ and $\tau$ from
the first geodesic equation in \eqref{eqn:4_vel_ex1}. Solving
\begin{equation}
    \frac{dt}{d\tau} = \sqrt{\frac{8 \pi C}{3t}},
    \label{eqn:geodesic_ex1}
\end{equation}
subject to the boundary condition $t=\tau=0$ at the big bang, one finds
\begin{equation}
    t={\left(6\pi C\tau^2\right)}^{1/3}.
    \label{eqn:t_ex1}
\end{equation}
We may verify that this relationship is correct by noting that it
leads to the appropriate expression for the density as a function of the
proper time of comoving observers in an EdS universe, namely
\begin{equation}
    \rho = \frac{1}{6\pi \tau^2}.
    \label{eqn:rho_in_tau_ex1}
\end{equation}
Hence, the coordinate $t$ here is simply proportional to
$\tau^{2/3}$. Note that the specific relation between $t$ and
$\tau^{2/3}$ was determined by our choice in $a(t)$; any other choice
for $a(t)$ would also yield constant spatial components in the
4-velocity, and the frame would be declared `comoving', but the $t$
would not (in general) be the proper time of the comoving observer in
that frame, and this occurs without any inconsistencies or
restrictions.

This one counter-example is sufficient to prove that a gauge
transformation of the form~\eqref{eqn:melia_gauge_transf} is allowed,
and $A$ does not necessarily have to be constant. When $A=1$, we are
in the frame of the comoving/freely-falling observer, with the
coordinate $t$ equal to their proper time. When $A=A(t)$, the
spatial coordinates still are those of the comoving/freely-falling
observer, but the coordinate $t$ is simply some function of their
proper time, and this function is determined by the specific form of
$A(t)$. In this case, one can simply use the gauge transformation
given by~\eqref{eqn:melia_gauge_transf} to bring us back to
the conventional FRW metric with $A=1$, in which the time coordinate is
equal to the proper time of the comoving observer.

Finally, we address a related part of Melia's argument
\citep{Melia2016a}, in which he claims that it is inconsistent with
basic relativity theory to have $A=1$ in a cosmological model with
$\ddot{a} \neq 0$. He bases this claim on the fact that one can always
distinguish between accelerated and inertial frames. In particular,
Melia suggests that the accelerated universal expansion should produce
a time dilation that is measurable relative to the passage of proper
time in the (inertial) free-fall frame, and hence $A$ cannot be unity. 
The flaw in Melia's argument is that the condition
$\ddot{a} \neq 0$ represents a coordinate acceleration rather than a
proper acceleration. Any observer comoving with the cosmological 
fluid follows a geodesic and is hence freely-falling and so does not
experience any proper acceleration, and this is perfectly consistent
with having $\ddot{a} \neq 0$.

 
\section{Conclusions}

To summarise, Melia's first claim is that for $A$ in
\eqref{eqn:metric_comov} to be a constant, one requires
$\rho+3p=0$. We have shown that this is false, and results simply from
a false step in logic, and we have provided an example using the EdS
cosmology that demonstrates this. Secondly, Melia claims that $A$ (and
hence $\Phi$ using Melia's own notation) needs to be constant, by
arguing that the free-fall and comoving frames must coincide. We have
shown that the two frames can coincide perfectly well even with $A$
not constant; in this case, the coordinate time $t$ is no longer the
proper time of comoving observers, but a function of it arising via a
simple gauge transformation.


Thus, contrary to the claims presented in~\cite{Melia2016}, there is
no extra information to be extracted from starting by substituting the
general spherically-symmetric metric into the Einstein equations and
then imposing homogeneity and isotropy, as compared to the usual route
of first imposing homogeneity and isotropy on the metric and then
employing the Einstein equations. Hence, the FRW spacetime is
perfectly compatible with having $\rho+3p \neq 0$.

In closing, it is also worth pointing out here that the many claims
that the $R_{\rm h}=ct$ model is favoured over $\Lambda$CDM by
observational data should also be treated with caution.  As discussed
in the Introduction, more recent observational data cast doubt on the
model's central assumptions, but there exists a further issue related
to how the $R_{\rm h}=ct$ and $\Lambda$CDM models have previously been
compared. In the $\Lambda$CDM model, there is no requirement that
$\rho+3p = 0$. This condition is, however, broadly consistent with
much of the observational data, as has been known for some time.
Thus, if one merely imposes this additional condition \emph{post-hoc}
on the $\Lambda$CDM model, to arrive at the $R_{\rm h}=ct$ model, then
any model selection approach will naturally favour the latter. Such
analyses have content only if one has a physical reason \emph{a
  priori} to impose the zero active mass condition. As we have shown,
the argument presented in~\cite{Melia2016} for imposing this condition
is not valid.

\section*{Acknowledgements}
Do Young Kim is supported by a Samsung Scholarship.

\bibliographystyle{mnras}
\bibliography{ref_w_abbrev}



%
%
%

\bsp
\label{lastpage}
\end{document}